\documentclass[aps,pra,floatfix]{revtex4}
\usepackage[dvips]{graphicx}
\usepackage[dvips,letterpaper,margin=30mm]{geometry}
\newcommand{\eps}{\epsilon}
\newcommand{\dagg}{\dagger}

\begin{document}
	
	\title{High-precision study of Cs polarizabilities}
	
	\author{E. Iskrenova-Tchoukova, M. S. Safronova}
	  
	\affiliation{Department of Physics and Astronomy, University of Delaware, Newark, 19716, USA}
	
	\author{ U. I. Safronova }
\email{usafrono@nd.edu}\altaffiliation{ On  leave  from ISAN,
Troitsk, Russia} \affiliation{Physics Department, University of
Nevada, Reno, NV
 89557}

	\begin{abstract}
We present results of the first-principles calculation of Cs dipole static polarizabilities for the $Ns$ ($N = 6 - 12$), 
$Np_j$ ($N = 6 - 10$), and $Nd_j$ ($N = 5 - 10$) states using the relativistic all-order method. In our implementation 
of the all-order method, single and double excitations of Dirac-Fock wave functions are included to all orders 
in perturbation theory. Additional calculations are carried out for the dominant terms and the uncertainties of our
final values are estimated for all states. A comprehensive review of the existing theoretical and experimental 
studies of the Cs polarizabilities is also carried out. Our results are compared with other 
values where they are available. These calculations provide a theoretical benchmark for a large number of Cs polarizabilities.
	\end{abstract}

	\date{\today}
	
	\maketitle

\section{Introduction} 

Atomic and molecular polarizabilities are of great interest because various properties of matter 
can be expressed in terms of multipole moments and polarizabilities of the atoms or molecules in the system. 
Polarizabilities describe the response of a system to external electric fields.
Therefore, atomic polarizabilities reflect the atomic structure and can be used to probe correlation and relativistic 
effects. On the other hand, atomic polarizabilities determine the long-range van der Waals 
interactions between the atoms and are used in describing atomic  scattering processes. Atomic polarizabilities 
play an important role in high Rydberg state spectroscopy (e.g. \cite{snow:2005}). 
The study of the alkali-metal atoms are of particular interest because they allow for very accurate
comparison between the experiment and theory. The conclusions reached from such studies may also be useful for the 
understanding of more complicated systems. 
Cs is also of particular interest owing to the study of the 
parity nonconservation (PNC), designed to test the standard model of 
the electroweak interaction and to
set limits on its possible extensions as well as to infer nuclear anapole moments.
 The highest accuracy of such an experiment was reached
 for cesium, where measurements of PNC amplitudes have reached an accuracy of 0.4\% 
\cite{wood:1997}. To make meaningful tests of the standard model, high-precision calculations
of the PNC amplitudes must be carried out at a similar
level of accuracy, and the uncertainty of the  calculation has to be established. 
 
In this work, we carry out  a systematic study of a large number of Cs polarizabilites 
in order to provide recommended values for the 
$Ns$ ($N = 6 - 12$), 
$Np_j$ ($N = 6 - 10$), and $Nd_j$ ($N = 5 - 10$) states and evaluate their uncertainties. 
The best-set values for the 91 electric-dipole matrix elements
used in our calculations are also provided with their uncertainties. These data are also 
useful for a number of other applications. 

\subsection{Experimental methods and studies of the atomic polarizabilities}

In this section, we provide a summary of a variety of methods used to measure the 
atomic polarizabilities as well as describe the development in the experimental measurements of the 
electric-dipole polarizability of cesium ground state.

In 2005, Gould and Miller \cite{gould:2005} wrote a comprehensive review of the 
experimental methods to determine the static electric-dipole polarizabilities.
 Miller and Bederson's earlier review from 1988 \cite{miller:1988} 
concentrated on the bulk polarizability measurements and the atomic beam methods. 
Average bulk ground state static polarizabilities are measured by determining the dielectric constant of an atomic 
or molecular gas. The bulk dynamic polarizabilities are determined by measuring the refractive index 
of the gas, see \cite{miller:1988}. The bulk methods are very accurate, but their limitation lies in the need to deal with atoms or molecules that are stable and gaseous at room temperature
and the fact that the effect of the excited states can not be accounted for.

In 1974, Molof {\it et al.} \cite{molof:1974} used the E-H-gradient balance technique to measure the static electric-dipole polarizabilities of alkali-metal atoms. They obtained the value (59.6 $\pm$ 1.2)~$10^{-24}$ cm$^3$ for electric-dipole polarizability of the ground state of cesium. 
Hall and Zorn \cite{hall:1974} measured the value (63.3 $\pm$ 4.6)~$10^{-24}$ cm$^3$ for the  electric-dipole polarizability 
of the ground state of cesium. They used the deflection of a velocity-selected atomic beam in 
inhomogeneous electric field. The technique is based on the fact that the deflection experienced by atoms moving 
through a region with known transverse electric field gradient is proportional to the dipole polarizability of the atoms.
An important detail of this technique is that the precision with which the velocity of the atoms is known 
puts a limitation on the precision of the experiment. The short interaction time 
in the case of high velocity which leads to small deflection of the beam places 
another limitation on the accuracy of this method.

In 1995, Ekstrom {\it et al.} \cite{ekstrom:1995} designed an atomic interference experiment that allowed them to measure the ground state energy shifts with spectroscopic precision and determine the 
ground state dipole polarizability. In 2003, Amini and Gould \cite{amini:2003} designed an experiment that avoids the problems associated with the 
measuring the deflection of a thermal beam in transverse electric-field gradient. 
They measure the effect of the electric-field gradient on the longitudinal velocity of the beam 
of cesium atoms in  a magneto-optical trap (MOT).  
The cesium $6s$ scalar dipole polarizability is found from the time-of-flight of laser cooled and 
launched cesium atoms traveling through an electric field. The cited value is
(59.42 $\pm$ 0.08)~10$^{-24}$ cm$^3$. This is the most precise measurement of the ground state 
polarizability at this time. 

Another group of experiments allows to infer the atomic polarizabilities by measuring the Stark shift of the cesium lines, 
e.g. \cite{Stark1}. In 1988, Tanner and Wieman \cite{tanner:1988} measured the Stark shift 
in the 6s$_{1/2} \longrightarrow 6{\rm P}_{3/2}$ transition in Cs. The dc Stark shift of the cesium D1 line has been has been measured to 0.01\% in Ref.~\cite{Stark2}. The authors of this work \cite{Stark2} noted that it was the most precise Stark shift measurement ever reported. 
The Stark shifts of the $6p_{3/2}-(10-13)s$ states in cesium were measured in Ref.~\cite{WW}.
The Stark shifts of cesium 11$D$ states were measured with high precision by van Wijngaarden and Li 
in 1997 \cite{wijngaarden:1997} using an electro-optically modulated laser beam.  The authors note that the tensor polarizabilities
reported in that work \cite{wijngaarden:1997} were the most accurate yet determined for any atomic state. 
The dc Stark shift of the $6s-7s$ transition in atomic cesium was measured with high precision in 1999 \cite{new7s}
using laser spectroscopy. The result of this experiment  disagrees with a previous measurement \cite{watts:1983}
 but was within 0.3\% of the value predicted by the \textit{ab initio} calculations \cite{dzuba,blundell:92}
   removing the largest at that time 
 outstanding disagreement between experiment and \textit{ab initio} theory of low-lying states in atomic cesium.

The atomic polarizabilities can be derived from measurements of the lifetimes 
of the corresponding levels. The contribution of the core electrons to the polarizability of the alkali atoms is
 small. Then, the main contribution to the ground s-state polarizability comes from the first 
low-lying excited P-states, i.e. dominant electric-dipole matrix elements are 
$\langle Ns | {\mathbf D} | np_{1/2, 3/2} \rangle $; 
see \cite{safronova:2004,derevianko:2002} for a detailed discussion and comparison of 
lifetime and polarizability measurements in cesium.

A large group of experiments makes use of the level-crossing of some hyperfine atomic levels at 
finite electric field. The first observation of the purely electric field level-crossing 
was reported in 1966 \cite{khadjavi:1966}. This type of measurements allows for experimental
determination of the excited states tensor polarizabilities. Recent cesium  measurements 
were reported by  Auzinsh {\it et al.} \cite{auzinsh:2006,auzinsh:2007}.

\subsection{Theoretical studies of cesium polarizabilities}

Since the alkali-metal atoms are monovalent systems, they represent an excellent opportunity to 
study the correlation effects. Heavy alkalis are of particular interest owing to the application 
to the study of fundamental symmetries. The polarizability of the alkali-metal atoms
are essentially the same as the valence polarizability as the contribution of the ionic 
core was determined to be small \cite{safronova:1999,derevianko:1999}. We summarize the theoretical 
studies of Cs polarizabilities below. 

In his seminal paper \cite{dalgarno:1962}, Dalgarno summarized the mathematical foundation of the theory of 
the atomic perturbation and discussed the methods of calculating the atomic polarizabilities and shielding factors. 
The polarizability of the cesium atom obtained by Dalgarno and Kingston \cite{dalgarno:1959} using the 
oscillator-strength formula was (53.7 $\pm$ 5.4)~ $10^{-24}$~cm$^3$.

According to the oscillator-strength formula, the knowledge of the (reduced) electric-dipole matrix elements 
is crucial for calculation of the atomic polarizabilities. The reduced matrix elements can be 
computed in a number of approximations. Variety of theoretical methods are used, such as 
 third-order many-body perturbation theory, multiconfiguration Hartree-Fock (MCHF),
configuration interaction (CI) method, coupled-cluster (CC) method, and it relativistic linearized 
version referred to as the all-order method as well as others.  

In 1970, Sternheimer \cite{sternheimer:1970} used the Hartree-Fock wave functions to compute the
quadrupole polarizability of some ions and alkali atoms. The cesium ground state value was calculated to be
71.31 $ 10^{-24}$ cm$^3$. 
In 1971, Schmieder {\it et al.} \cite{schmieder:1971} calculated the scalar and quadrupole 
polarizabilities of cesium $p_{3/2}$ states in the second order perturbation theory. 
The work by Kell\"o {\it et al.} \cite{kello:1993} contains a detailed investigation of the contracted 
Gaussian basis sets in the calculation of the electric-dipole polarizabilities of alkali-metal atoms. 
The calculations are performed using the complete-active-space self-consistent field and second 
order perturbation theory, CASSCF and CASPT2. 
Another group of Gaussian basis set methods  use relativistic pseudopotentials (see \cite{lim:2005} 
and the references there). Pseudopotential methods replace the core electrons by an effective, 
pseudopotential. The core polarization and the relativistic effects are incorporated as well.
The Douglas-Kroll relativistic CCSD(T) method with the optimal basis set gives 
58.09~10$^{-24}$ cm$^3$ for the cesium ground state dipole polarizability.

Extensive calculation of the polarizabilities of 
cesium $Ns$, $Np$, $Nd$, and $nF$ states  was carried out by van Wijngaarden and Li \cite{wijngaarden:1994}   using the Coulomb approximation. They also provided extensive comparison with other theoretical and experimental values. 

Patil and Tang \cite{patil:1997} computed the multipolar polarizabilities, $\alpha_q$, with $q=1,2,\dots,12$,
for the alkali isoelectronic sequences. The ground state wave functions were taken to be the asymptotically 
correct wave functions, i.e. the two leading terms in the asymptotic expansion of the wave function are retained. 
The excited states are taken to be the Coulomb wave functions with a correction that makes sure the 
experimental energies of the low-lying states are reproduced correctly. The ground state electric-dipole 
polarizability of cesium was found to be 60.6~10$^{-24}$ cm$^3$.

The relativistic linearized version of the coupled-cluster 
singles-doubles method, i.e. all-order SD method, was used in \cite{safronova:1999,derevianko:1999}
 to calculate the static dipole polarizabilities of the 
alkali-metal atoms. This method is discussed in more details in Section \ref{compmethod}.
The value obtained for the ground state static dipole polarizability is 59.3(3)~10$^{-24}$ cm$^3$   \cite{derevianko:1999}.
In \cite{porsev:2003}, Porsev and Derevianko computed the ground state 
quadrupole and octupole polarizabilities
of the alkali-metal atoms using the relativistic  MBPT. 

In 2004, Safronova and Clark \cite{safronova:2004} 
pointed out the inconsistencies between the lifetime and polarizability measurements in cesium.
The dominant contribution to the $6p$ scalar polarizability comes from the $5d-6p$ matrix elements.
This allows for a check of the accuracy of the matrix elements. The authors compare the values of the 
$6p$ polarizabilities obtained using the relativistic all-order SD method and using the values of the
matrix elements derived from the $5d$ lifetime experiment \cite{5dtau}. They point out that the theoretical 
all-order values yield a polarizability result in agreement with the polarizability measurements \cite{Stark2,tanner:1988} but not
 with the lifetime measurements \cite{5dtau}.

In a recent work, Gunawardena {\it et al.} \cite{mevan:2007} report a precise measurement of the dc Stark shift 
of the $6s \longrightarrow 8s$ transition in atomic cesium. The experiment makes use of the 
Doppler-free two-photon absorption measurement. The value of the 
static polarizability of $8s$ state in cesium, extracted from the experiment, is 38 060$\pm$250 $a_0^3$.
The authors present a theoretical value of 38 260$\pm$290 $a_0^3$. The theoretical value of the polarizability 
is calculated using the relativistic all-order SD method. 

\section{Method}

The energy shift of the  $|\gamma j m\rangle$ atomic level in a dc electric field $\bf{E}={\cal E}\hat{\bf{z}}$
is given by  

\begin{equation}\label{pol1}
\Delta E  = - \frac{1}{2} \alpha_{\gamma j m} {\cal E}^2,
\end{equation}

\noindent where $\alpha_{\gamma j m}$ defines the static polarizability of the corresponding atomic state 
$|\gamma j m\rangle$. The scalar and tensor static polarizabilities $\alpha_{0, \gamma j m_j}$ 
and $\alpha_{2, \gamma j m}$ are defined as 
\begin{equation}\label{scal_tensor}
\alpha_{\gamma j m} = \alpha_{0,{\gamma j m}} + \alpha_{2,{\gamma j m}} \frac{3m^2 -j(j+1)}{j(2j-1)}.
\end{equation}
We omit index $\gamma j m$ in the text below and refer to scalar and tensor static polarizabilities as 
$\alpha_0$ and $\alpha_2$, respectively. 

We separate the calculation of the scalar static polarizability into the calculation of the 
polarizability of the ionic core and the valence polarizability. The random-phase-approximation (RPA) calculation of the Cs core polarizability was carried out in Ref.~\cite{johnson:83} and yielded the value $15.8 a_0^3$, where $a_0$ is the Bohr radius. Based on the evaluation of the accuracy of RPA
approximation for the polarizabilities of the noble gases, this value is accurate to at least 5\%. 
The core polarizability is small even in comparison with the valence polarizabilities of the 
lowest states such as $6p$ and $5d$. It is negligible for the higher states.  For example, core
polarizability contributes only 4\% to the total value of the ground state polarizability and only 1\% to the $6p_{1/2}$
polarizability. Therefore, the RPA value of the core polarizability 
is sufficiently accurate for this work. The separation of the scalar polarizability
to the core and valence parts also produces a  compensation term that accounts for the 
Pauli exclusion principle, i.e. in Cs it subtracts 1/2 of the core polarizability contribution associated with the 
excitation to the valence shell. This term is only 2\% of the core contribution \cite{safronova:1999} even for the ground state and, therefore, below the estimated uncertainty of the core term itself. It is  negligible for all other states.   

The valence scalar and tensor static polarizabilities of the atomic state $|\gamma j \rangle$ are expressed in lowest order 
as sums over unperturbed intermediate states $|\beta j_{\beta} \rangle$ of parity opposite to that of the state 
$|\gamma j \rangle$:

\begin{eqnarray}\label{scal}
\alpha_{0} & = & \frac{2}{3(2j+1)} \sum\limits_\beta \frac{|\langle \gamma j || D || \beta j_{\beta}  \rangle |^2}{E_{\beta} - E_\gamma},
\end{eqnarray}

\begin{eqnarray}\label{tensor}
\alpha_{2} & = & 4 \sqrt { \frac{5j(2j-1)}{6(2j+3)(2j+1)(j+1)}}
                  \sum\limits_{\beta} (-1)^{j+j_{\beta}} \left\{ 
                                                 \begin{array}{ccc}
                                                   j & 1 & j_{\beta} \\
                                                   1 & j & 2 \\
                                                 \end{array} \right\} \frac{|\langle \gamma j || D || \beta j_{\beta} \rangle |^2}{E_{\beta} - E_{\gamma}},
\end{eqnarray}
where $\langle \gamma j || D || \beta j_{\beta} \rangle$ is the reduced electric-dipole matrix element defined as
\begin{equation}\label{reduced}
\langle \gamma j m |  D_q | \gamma^{\prime} j^{\prime} m^{\prime} \rangle = 
(-1)^{j^{\prime}-m^{\prime}} 
\left( \begin{array}{ccc}
                                                            j^{\prime} & 1 & j \\
                                                           -m^{\prime} & q & m \\
                                                           \end{array} \right) 
\langle \gamma j || D || \gamma^{\prime} j^{\prime}  \rangle,
\end{equation}
and the $D_q$ is the corresponding component of the electric-dipole operator in 
spherical coordinates.

The sums over states $\beta$ in Eqs.~(\ref{scal},\ref{tensor}) separate into the 
two or three sums over the principal quantum number for each type of the 
allowed electric-dipole transitions for Cs calculation. The allowed values of $\beta$ are the following:  $\beta = np_{1/2}, np_{3/2}$ for 
$Ns$ states, $\beta = ns, nd_{3/2}$ for the $np_{1/2}$ states, $\beta = ns, nd_{3/2}, nd_{5/2}$
for the $np_{3/2}$ states,  
$\beta = np_{1/2}, np_{3/2}, nf_{5/2}$ for the $nd_{3/2}$ states, and  $\beta = np_{3/2}, nf_{5/2}, nf_{7/2}$
for the $nd_{5/2}$ states. here, $n$ and $N$ are principal quantum numbers. 
We note that there is no tensor contribution to the polarizability of the $Ns$ and $Np_{1/2}$
states. 

In order to evaluate the sums over the principle quantum numbers $n$, 
we carry out all calculations in a finite B-spline basis set \cite{johnson:88}  constrained to a large 
spherical cavity and defined on a non-linear grid. Therefore, the sums in Eqs.~(\ref{scal},\ref{tensor})
range over the basis set states. In order to study such highly-excited states as $12s$ and $10d$, we needed to use
a very large cavity, $R = 220$~a.u, to ensure that the corresponding wave functions fit inside the 
cavity. As a result,  we had to use a large number of basis set functions, $N_B=70$, in order to correctly 
represent the properties of both highly-excited states and the lowest states. We verified that the basis 
set energies match the Dirac-Fock energies for all of the states considered in this work.
The order of splines was taken to be $k=8$. More calculation details  associated with the 
use of the finite basis set are described in Section~\ref{compmethod}. 

The sums over the principal quantum numbers in Eqs.~(\ref{scal},\ref{tensor}) 
converge very rapidly, with the exception of the 
sums involving the $5d_{3/2}-nf_{5/2}$ and $5d_{5/2}-f_{7/2}$ transitions 
which we will discuss separately in Section\ref{results}. In fact, each of the sum over the principal quantum 
number is dominated by one or two terms that correspond to the lowest possible values of the 
denominator $E_{\beta} - E_\gamma$ in Eqs.~(\ref{scal},\ref{tensor}). A small number of other terms may also
be significant for the precise calculation for some states. As a result, only a few terms from  each sum have to be calculated accurately, allowing us to separate the 
calculation of the valence scalar and tensor polarizabilities into the two parts, the main term containing 
all significant contributions and the tail:
\begin{equation}
 \alpha_{0, 2} = \alpha_{0, 2}^{\text{main}} + \alpha_{0, 2}^{\text{tail}}.  
 \end{equation}  
The separation of $\alpha$ into the main and tail parts is done independently for each of the two or three
sums over the principal quantum number $n$ contributing to the polarizability of the specific state:
\begin{equation}
 \sum_n = \sum_{n_0}^{n_{\text{main}}} + \sum_{n_{\text{main}}+1}^{N_B},
 \end{equation}  
where the $n_0$ is the lowest possible value of principal quantum number for the particular 
set of $\beta$ states, $n_{\text{main}}$ is the cut-off that we select for the separation of the 
main and tail terms, and $N_B$ is the number of the finite basis set orbitals set to 70 in the 
present work.  In general, $n_{\text{main}}$ may be selected differently for the specific state $\gamma$
and each $\beta$ sum, but we chose to use the same  $n_{\text{main}}$ for 
all of the states considered in this work. 
We use $n_{\text{main}} = 12$ for $\beta={ns, np_{1/2}, np_{3/2}}$, $n_{\text{main}} = 10$ for  
$\beta = {nd_{1/2}, nd_{3/2}}$, and $n_{\text{main}} = 8$ for  
$\beta = {nf_{5/2}, nd_{7/2}}$, respectively. The only exception is the addition of the $9f$ contributions
to the calculation of the $10d$ polarizabilites. 

Such high value of the cut-off principal quantum number also reduced the tail contribution and improved 
the accuracy of our calculations. The remaining tail contributions are 
evaluated using in the Dirac-Fock (DF) approximation, i.e. both energies and 
E1 matrix elements were calculated in the DF approximation. 

In summary, we reduce the calculation of the polarizabilities to the calculation of the 
electric-dipole reduced matrix elements required for the evaluation of the main terms for each state. 
We use the experimental energies from Refs.~\cite{SW, NIST1, NIST2} in the calculation of the 
main terms. 
Owing to the large number of states considered in this work,  
317 transitions contribute to the main term and 102 transitions give dominant contributions. We calculated all of the  
317 electric-dipole matrix elements using the relativistic all-order method and conducted additional 
calculation for the 102 transitions that involved the evaluation of the largest missing corrections 
and evaluation of the uncertainty of the final values. The calculation of the matrix elements is 
described in the next section.   

\section{Calculation of the E1 matrix elements}\label{compmethod}

We carry out the calculation of the electric-dipole reduced matrix elements using the relativistic SD all-order method where single and double excitations of the (frozen-core) Dirac-Fock wave function
are included to all orders in perturbation theory \cite{blundell:89,safronova:1999}.
Triple excitations are also partially included for selected cases. 
The relativistic SD all-order method is a linearized coupled-cluster method restricted to 
single and double excitations.
A comprehensive review of the coupled-cluster method and its applications in quantum chemistry
is given in Ref.~\cite{CS:02}.
In the coupled-cluster method, the exact wave function of the monovalent atom in a state
 $v$ is represented as
\begin{equation}
|\Psi_v\rangle = \exp{(S)} |\Phi_v \rangle, \label{eqtt}
\end{equation}
where $\Phi_v$ is the lowest-order atomic wave function for the state $v$, which
taken to be a frozen core Dirac-Fock (DF) wave function in our calculations. 
The cluster operator $S$ is expressed as a sum of $n$-particle excitations $S_n$ of the 
lowest-order wave function
\begin{equation}
S=S_{1} + S_{2} + S_{3} + \cdots.
\end{equation}

The exponential function in Eq.~(\ref{eqtt}) can be expanded to give
\begin{equation}
|\Psi_v \rangle=(1+S+\frac{1}{2}S^2+\cdots) |\Phi_v\rangle.
\label{eqq}
\end{equation}

In the linearized single-double (SD) coupled-cluster method, only terms that are linear in the $S_i$
remain and all other terms, for example $S_1 \times S_2$ are omitted , i.e. 
\begin{equation}
|\Psi_v\rangle = \left(1+S_1+S_2+ \cdots \right) |\Phi_v \rangle. \label{eq1}
\end{equation}

The contributions from the non-linear terms were recently investigated in
Refs.~\cite{Andrei1, Andrei2, usNL}. We refer the reader to Ref.~\cite{usNL} 
for a complete list of the non-linear terms and detailed investigation of their
contributions to the alkali-metal atom properties. 
The computational complexity of the calculations increases dramatically beyond
the double excitations term $S_2$, and we include triple excitations partially in some 
of the calculations using a perturbative approach. 
We note that in this work very large ($N_B=70$ for each partial wave) 
complete basis set is used to correctly 
reproduce necessary atomic properties for highly excited atomic states, requiring
significant computational resources for the SD all-order calculations.

The expression for the single excitations is given by 
\begin{equation}
S_{1} = \sum_{ma} \rho_{ma} a^{\dag}_m a_a + \sum_{m\neq v} \rho_{mv} a^{\dag}_m a_v,
\end{equation}
where the first term corresponds to single core excitations and the second term
corresponds to single valence excitations. The expansion coefficients $\rho_{ma}$ 
and $\rho_{mv}$  are referred to as
single core and valence excitation coefficients, and the 
 $a^\dagger_i$ and $a_i$ are creation and
annihilation operators for an electron in a state $i$. We use the letters from the beginning
of the alphabet $a, b, ...$ to designate core orbitals and letters from the 
middle of the alphabet, $ m, n, ...$ to designate excited states. For Cs, we 
include all 17 $a = 1s, ..., 5p_{3/2}$ core shells in our calculations. 

The double excitation term is given by 
\begin{equation}
S_{2} = \frac{1}{2}\sum_{mnab}\rho_{mnab} a^{\dag}_m a^{\dag}_n a_b a_a +
\sum_{mnb}\rho_{mnvb} a^{\dag}_m a^{\dag}_n a_b a_v,
\end{equation}
and the quantities $\rho_{mnab}$ and $\rho_{mnva}$ are
referred to as double core and valence excitation coefficients, respectively.

Therefore, the atomic wave function in the all-order SD method  \cite{blundell:89} 
is expressed via the single and double excitation coefficients as follows:
\begin{eqnarray} \nonumber
&& |\Psi_v^{\rm SD} \rangle  =  \left( 1 + \sum\limits_{ma}^{} \rho_{ma} a_{m}^{\dagg} a_{a} + \frac{1}{2} \sum\limits_{mna
b}^{} \rho_{mnab} a_{m}^{\dagg} a_{n}^{\dagg} a_{b} a_{a} \right. \\ \label{wave}
       &&  +   \left. \sum\limits_{m \neq v} \rho_{mv} a_{m}^{\dagg} a_v
        +  \sum\limits_{mna} \rho_{mnva} a_{m}^{\dagg} a_{n}^{\dagg} a_a a_v \right)|\Phi_v\rangle .
\end{eqnarray}

The equations for the excitations coefficients and the correlation energy 
are derived by substituting the SD all-order wave function given by the Eq.~(\ref{wave}) 
into the  Schr\"{o}dinger equation
  \begin{equation}
H | \Psi_v\rangle=E| \Psi_v\rangle, \label{eq2}
\end{equation}
where the Hamiltonian $H$ is the relativistic {\em
no-pair} Hamiltonian \cite{nopair}, which can be
 written in second-quantized form as
\begin{equation}
H = \sum\limits_{i}^{} \eps_{i}  a_{i}^{\dagg} a_{i}
+ \frac{1}{2} \sum\limits_{ijkl}^{} g_{ijkl}  a_{i}^{\dagg} a_{j}^{\dagg} a_{l} a_{k}
- \sum\limits_{ij} U_{ij}  a_{i}^{\dagg} a_{j},
\end{equation}
where $\eps_{i}$ are the one-body DF
energies for the state $i$, $ U_{ij}$ is taken to be frozen-core 
DF potential in our calcuatlion, and $g_{ijkl}$ are the two-body Coulomb integrals:
\begin{equation}
g_{ijkl} = \int \, d^3 r \int \, d^3 r' \, \psi^{\dagg}_i ({\bf r}) \psi^{\dagg}_j ({\bf r'})
\frac{1}{|{\bf r} - {\bf r'}|} \psi_k ({\bf r}) \psi_l ({\bf r'}).
\end{equation}

For example, the equation for the single valence excitation
coefficients $\rho_{mv}$ is given  by
\begin{eqnarray}
(\epsilon_v-\epsilon_m + \delta E_v)\rho_{mv} &= & \sum_{bn}
\tilde{g}_{mbvn}\rho_{nb}  + \sum_{bnr}
g_{mbnr}\tilde{\rho}_{nrvb} - \sum_{bcn}
g_{bcvn}\tilde{\rho}_{mnbc}, \label{v1}
\end{eqnarray}
where $\delta E_v$ is the correlation correction to the valence
energy for the state $v$ given in terms of the
excitation coefficients by
\begin{equation}
\delta E_v = \sum_{ma} \tilde{g}_{vavm}\rho_{ma} + \sum_{mab}
g_{abvm}\tilde{\rho}_{mvab} + \sum_{mna}
g_{vbmn}\tilde{\rho}_{mnvb}. \label{env}
\end{equation}
We use the designation 
 $\tilde{\rho}_{mnab}=\rho_{mnab}-\rho_{mnba}$ and $\tilde{g}_{mnab}=g_{mnab}-g_{mnba}$
 in the equations above.
The sum over the magnetic quantum numbers is carried out analytically and the resulting
 equations are solved iteratively for the excitation coefficients until the 
correlation energy converges. The excitation coefficients are then used for the calculation of
the matrix elements as described below.

In general, the one-body operator $Z$ can be written in second quantization as
$Z = \sum_{ij} z_{ij} a^\dagg_i a_j$.

\noindent The expression for SD matrix elements of operator $Z$ is obtained by substituting
the SD wave function given by Eq.~(\ref{wave})  into the expression 
\begin{equation}
Z_{wv}=\frac{\langle \Psi_w |Z|\Psi_v \rangle}{\sqrt{\langle \Psi_w| \Psi_w \rangle\langle \Psi_v| \Psi_v \rangle}}. \label{z}
\end{equation}
The resulting SD matrix element is given by  
\begin{equation}
Z_{wv}=\frac{z_{wv} + Z^{(a)} + \dots + Z^{(t)}}{\sqrt{(1+N_v)(1+N_w)}}, \label{zx}
\end{equation}
where $z_{wv}$ is the DF matrix element, terms  $Z^{(a)}, \dots Z^{(t)}$ are linear or 
quadratic functions of the excitation coefficients, and $N_v$ and $N_w$ are normalization terms that  
are quadratic functions of the excitation coefficients. 
For most of the dominant transitions in our polarizability calculations, a single term 
\begin{equation}
 Z^{(c)} = \sum_{m}z_{wm} \rho_{mv} +\sum_{m} z_{mv} \rho_{mw} 
\end{equation}
gives the dominant contribution. Two other terms, 
\begin{equation}
  Z^{(a)} = \sum_{ma}z_{am}\tilde{\rho}_{wmva} + \sum_{ma} z_{ma} \tilde{\rho}_{vmwa} 
\end{equation}
and
\begin{equation}
    Z^{(d)} =  \sum_{mn} z_{mn} \rho_{mw} \rho_{nv}    
    \end{equation}
may be dominant for selected important transitions.  We note that both $Z^{(c)}$ 
and $Z^{d}$ terms contain only single valence excitations coefficients. The complete expression for the 
matrix elements is given in Ref.~\cite{blundell:89}. 

All sums over the excited state in the formulas above range over the 
basis set states. We truncated  last five basis set orbitals for each partial wave
since their contributions is negligible, i.e. 65/70 orbitals are included for each 
partial wave. All partial waves are included up to $l_{max}=6$, and orbitals with $j=l+1/2$ and $j=l-1/2$
are considered separately since it is intrinsically relativistic calculation. The basis set 
is numerically stable, i.e. the increase of the number of the basis set orbitals 
does not change the results within the current accuracy. The 
numerical uncertainty associated with the truncation of the number of the partial waves at $l_{max}=6$
is also negligible. We estimated the contribution from higher partial waves to be 0.1\% for the $6s-6p_j$
 transitions. The evaluation  of the higher partial wave contribution 
is carried out by conducting the third-order perturbation theory calculation carried out as described 
in Ref.~\cite{ADNDT} with the
same basis set and with higher number of the partial waves. We also verified that the 
use of the very large cavity did not affect the numerical accuracy of the 
atomic properties of the  lower states by comparing the present results for the transitions
between the lower states with all-order calculation carried out with the small 
cavity appropriate for the lower states.  We note that large basis set size, $N_B=70$,
is necessary to reproduce the atomic properties correctly with such a large cavity. We found that 
 the accuracy of the $N_B=50$ B-spline basis set is not sufficient for such calculation.  

As noted above, we have identified that the correlation correction for most of the dominant transitions in our polarizability calculation is essentially determined by a single term, $Z^{(c)}$, that contains
only single valence excitations. This term mostly corresponds to the Brueckner orbital correction 
as classified in the Ref.~\cite{ADNDT}. It is established (\cite{blundell:92,usca,usrdpd,auzinsh:2007}  that it
 can be corrected by scaling the single excitation coefficients using the ratio of the ``experimental'' and theoretical correlation energies and redoing the matrix element calculation with modified excitation coefficients. The ``experimental''
 correlation energies are determined as the differences of the experimental data and our lowest-order DF values. 
 We carry out such scaling where appropriate and determine the uncertainty of our calculation of the 
 matrix elements as the difference between the \textit{ab initio} and scaled data. In certain cases where this 
 correction is particularly large, we also
 carried out \textit{ab initio} calculation of the limited triple excitations and conducted
 more accurate study of the uncertainty of the calculations. The limited inclusion of the triples was also 
 aimed at correcting the $\rho_{mv}$ excitation coefficients. Such calculations are described in detail 
 in Refs.~\cite{usca,usrdpd,auzinsh:2007} and references therein.  We note that term $Z^{(d)}$ is also
 corrected by scaling as it  contains only valence single excitation coefficients. We note that the scaling procedure
  allows to 
 place an uncertainty on our theoretical data that is not derived from the comparison with the experiment. 
Our results are summarized in the next section. 
  
\section{Results and Discussion}\label{results}

 \begin{table*}
\caption{\label{me} Absolute values of the selected reduced electric-dipole matrix elements E1 in Cs and 
estimates of their uncertainties. Unless otherwise noted, these are all-order SD scaled values, including values from  Refs.~\cite{auzinsh:2007,mevan:2007}.  $^a$Experimental values, Ref.~\cite{berry}, $^b$SD all-order scaled values, previously published in Ref.~\cite{mevan:2007}, $^c$experimental values from Ref.~\cite{usbeta}, $^d$derived from the $7s-6s$ Stark shift value in Ref.~\cite{safronova:1999}, $^e$all-order values, Ref.~\cite{safronova:2004},
$^f$SD all-order scaled values, previously published in Ref.~\cite{auzinsh:2007}. Units: $ea_0$. }
\begin{ruledtabular}
\begin{tabular}{lrlrlr}
\multicolumn{1}{c}{Transition}&
\multicolumn{1}{c}{E1}&
\multicolumn{1}{c}{Transition}&
\multicolumn{1}{c}{E1}&
\multicolumn{1}{c}{Transition}&
\multicolumn{1}{c}{E1}\\
\hline \\[-0.4pc]
$6s - 6p_{1/2}$  &   4.489(7)$^a$&  $8s - 7p_{1/2}$  &   9.313(65)$^b$& $10s -  9p_{1/2}$ &   24.50(10)\\  
$6s - 7p_{1/2}$  &   0.276(2)$^c$&  $8s - 8p_{1/2}$  &   17.78(7) $^b$& $10s - 10p_{1/2}$ &   38.31(10)\\ 
$6s - 6p_{3/2}$  &   6.324(7)$^a$&  $8s - 7p_{3/2}$  &   14.07(7) $^b$& $10s -  9p_{3/2}$ &   36.69(10)\\  
$6s - 7p_{3/2}$  &   0.586(5)$^c$&  $8s - 8p_{3/2}$  &   24.56(10)$^b$& $10s - 10p_{3/2}$ &   52.67(16)\\ [0.5pc]
$7s - 6p_{1/2}$  &   4.236(21)$^c$&  $9s - 8p_{1/2}$   &  16.06(8) & $11s - 10p_{1/2}$  &  34.64(12)\\   
$7s - 7p_{1/2}$  &  10.308(10)$^d$&  $9s - 9p_{1/2}$   &  27.10(8) & $11s - 11p_{1/2}$  &  51.42(11)\\   
$7s - 6p_{3/2}$  &   6.473(32)$^c$&  $9s - 8p_{3/2}$   &  24.12(8) & $11s - 10p_{3/2}$  &  51.77(12)\\   
$7s - 7p_{3/2}$  &  14.320(14)$^d$&  $9s - 9p_{3/2}$   &  37.33(13)& $11s - 11p_{3/2}$  &  70.58(19)\\ [0.5pc]
$12s - 11p_{1/2}$  & 46.49(15)&$5d_{3/2} - 6p_{1/2}$&  7.06(16)$^e$& $5d_{5/2} - 6p_{3/2}$& 9.66(20)$^e$\\  
$12s - 12p_{1/2}$  & 66.43(13)&$5d_{3/2} - 6p_{3/2}$&  3.19(8)$^e$ & $5d_{5/2} - 4f_{5/2}$& 1.93(30)\\  
$12s - 11p_{3/2}$  & 69.37(15)&$5d_{3/2} - 4f_{5/2}$&  7.1(5)& $5d_{5/2} - 4f_{7/2}$& 8.6(6)\\
$12s - 12p_{3/2}$  & 91.1(2)  &                     &          &                      &         \\[0.5pc]
$6d_{3/2} - 6p_{1/2}$&  4.15(20)$^e$&  $6d_{5/2} - 6p_{3/2}$&  6.01(26)$^e$&$7d_{3/2} - 7p_{1/2}$&  6.56(2)$^f$\\   
$6d_{3/2} - 7p_{1/2}$&  18.0(2) &  $6d_{5/2} - 7p_{3/2}$&  24.4(3) &$7d_{3/2} - 8p_{1/2}$&  32.0(2)$^f$\\   
$6d_{3/2} - 6p_{3/2}$&  2.05(9)$^e$ &  $6d_{5/2} - 4f_{5/2}$&  6.60(5) &$7d_{3/2} - 7p_{3/2}$&  3.32(2)$^f$\\   
$6d_{3/2} - 7p_{3/2}$&  8.07(11)&  $6d_{5/2} - 5f_{5/2}$&  1.11(15)&$7d_{3/2} - 8p_{3/2}$&  14.35(8)$^f$\\  
$6d_{3/2} - 4f_{5/2}$&  24.6(2) &  $6d_{5/2} - 4f_{7/2}$&  29.5(2) &$7d_{3/2} - 4f_{5/2}$&  13.0(2)$^f$\\   
$6d_{3/2} - 5f_{5/2}$&  3.9(6)  &  $6d_{5/2} - 5f_{7/2}$&  4.96(67)&$7d_{3/2} - 5f_{5/2}$&  43.4(3)$^f$\\ [0.5pc]
$7d_{5/2} -  7p_{3/2}$&  9.64(4)$^f$ &  $8d_{3/2} -  8p_{1/2}$&9.18(5) & $ 8d_{5/2} -  8p_{3/2}$&  13.65(7) \\ 
$7d_{5/2} -  8p_{3/2}$&  43.2(2)$^f$ &  $8d_{3/2} -  9p_{1/2}$&49.3(2) & $ 8d_{5/2} -  9p_{3/2}$&  66.6(2) \\  
$7d_{5/2} -  5f_{5/2}$&  11.66(7)$^f$&  $8d_{3/2} -  8p_{3/2}$&4.71(01)& $ 8d_{5/2} -  5f_{5/2}$&  6.85(4) \\   
$7d_{5/2} -  4f_{7/2}$&  15.3(2)$^f$ &  $8d_{3/2} -  9p_{3/2}$&22.13(7)& $ 8d_{5/2} -  6f_{5/2}$&  17.54(8)\\  
$7d_{5/2} -  5f_{7/2}$&  52.2(3)$^f$ &  $8d_{3/2} -  5f_{5/2}$&26.1(2) & $ 8d_{5/2} -  5f_{7/2}$&  30.6(2) \\  
$                    $&          &  $8d_{3/2} -  6f_{5/2}$&65.2(4) & $ 8d_{5/2} -  6f_{7/2}$&  78.4(4)\\ [0.5pc] 
$9d_{3/2} -  9p_{1/2}$&  12.2(2) &  $9d_{5/2} -  9p_{3/2}$ & 18.3(2)  &$10d_{3/2} - 10p_{1/2}$ & 15.6(2) \\ 
$9d_{3/2} - 10p_{1/2}$&  70.0(2) &  $9d_{5/2} - 10p_{3/2}$ & 94.5(2)  &$10d_{3/2} - 11p_{1/2}$ & 94.1(2)\\  
$9d_{3/2} -  9p_{3/2}$&  6.33(6) &  $9d_{5/2} -  7f_{5/2}$ & 24.36(9) &$10d_{3/2} - 10p_{3/2}$ &  8.16(7)\\  
$9d_{3/2} - 10p_{3/2}$&  31.45(8)&  $9d_{5/2} -  6f_{7/2}$ & 49.3(3)  &$10d_{3/2} - 11p_{3/2}$ & 42.30(9)\\   
$9d_{3/2} -  6f_{5/2}$&  42.0(4) &  $9d_{5/2} -  7f_{7/2}$ & 108.9(4) &$10d_{3/2} -  7f_{5/2}$ & 61.0(2) \\
$9d_{3/2} -  7f_{5/2}$&  90.5(4) &                        &            &$10d_{3/2} -  8f_{5/2}$ & 119.4(4)\\ [0.5pc]
$10d_{5/2}- 10p_{3/2}$ & 23.5(3) & $10d_{5/2} - 8f_{5/2}$& 32.2(1)   & $10d_{5/2} - 8f_{7/2}$ & 143.8(5) \\ 
$10d_{5/2} - 11p_{3/2}$ & 127.1(3)  & $10d_{5/2} -  7f_{7/2}$ & 71.7(3)   & &\\
\end{tabular}
\end{ruledtabular}
\end{table*}
 \begin{table*}
\caption{\label{tab9s}  The contributions to the
scalar polarizability for the $9s$ state in cesium. The 
corresponding energy differences
and the absolute values of the lowest-order $Z^{DF}$ and final all-order electric-dipole
 reduced matrix elements $Z^{SD}$ are also listed. The energy differences are given in cm$^{-1}$.
 Electric-dipole  matrix elements  are given in atomic units ($ea_0$), and 
 polarizabilities are given in 10$^3$~$a_0^3$, where $a_0$ is Bohr radius. }
\begin{ruledtabular}
\begin{tabular}{lrrrrr}
\multicolumn{1}{c}{Contribution}&
\multicolumn{1}{c}{$\beta$}&
\multicolumn{1}{c}{$Z^{DF}_{\beta,9s}$ }&
\multicolumn{1}{c}{$Z^{SD}_{\beta,9s}$ }&
 \multicolumn{1}{c}{$E_{\beta}-E_{8s}$ }&
\multicolumn{1}{c}{$\alpha_0(9s)$}  \vspace{0.1cm}\\
\hline \hline \\[-0.4pc]
   $\alpha^{\text{main}}(nP_{1/2})$ & $6p_{1/2}$ &  0.56   &    0.55  &  -15732&   0.00        \\
                                    & $7p_{1/2}$ &  2.04   &    1.96  &   -5145&  -0.05        \\
                                    & $8p_{1/2}$ & 16.30   &   16.06  &   -1209& -15.7(2)    \\
                                    & $9p_{1/2}$ & 28.17   &   27.10  &     726&  74.0(4)    \\
                                    & $10p_{1/2}$&  2.67   &    2.76  &    1816&   0.31        \\
                                    & $11p_{1/2}$&  1.01   &    1.08  &    2493&   0.03        \\
                                    & $12p_{1/2}$&  0.56   &    0.60  &    2942&   0.01        \\
 $\alpha^{\text{tail}}(nP_{1/2})$   &            &         &          &        &   0.01  \\[0.5pc]
  $\alpha^{\text{main}}(nP_{3/2})$  & $6p_{3/2}$ &  0.79   &    0.77  &  -15178&     0.00      \\
                                    & $7p_{3/2}$ &  2.86   &    2.73  &   -4964&    -0.11     \\
                                    & $8p_{3/2}$ & 24.31   &   24.12  &   -1119&   -38.0(3)   \\
                                    & $9p_{3/2}$ & 38.99   &   37.33  &     771&   132.3(9)   \\
                                    & $10p_{3/2}$&  4.43   &    4.61  &    1843&     0.85     \\
                                    & $11p_{3/2}$&  1.80   &    1.93  &    2510&     0.11     \\
                                    & $12p_{3/2}$&  1.04   &    1.13  &    2954&     0.03     \\
 $\alpha^{\text{tail}}(nP_{3/2})$   &            &         &           &       &     0.04  \\[0.5pc]
  Total                             &            &         &           &       &    153.7(1.0)\\
\end{tabular}
\end{ruledtabular}
\end{table*}

The results of the all-order calculation of the matrix elements are summarized in Table~\ref{me}. Owing to the very large number of the transitions involved in this calculation, we only listed the transitions that give dominant contributions to the polarizabilities of the states considered in this work. In order to provide a 
best set of known data for these transitions, we replaced all-order theoretical values by the experimental ones where high-precision values are available. The experimental values from Ref.~\cite{berry} are used for 
the $6s-6p_j$ transitions, the values for the $6p_j-7s$ transitions are derived from the $7s$ lifetime 
measurement in \cite{usbeta}, and  the $6s-7p_j$ values are experimental values from \cite{usbeta}. The $7s-7p_j$ 
values were derived from the $7s-6s$ Stark shift measurement \cite{new7s}. We are quoting these $7s-7p_j$ values in the 
present table as the most accurate values available, and we used them in the calculation of the $7p_j$
polarizabilities to provide recommended values for these states. However, we used our theoretical values in the 
calculation of the $7s$ polarizabilites for the evaluation of the accuracy of our
calculation. Otherwise, the comparison of the $7s$ values with the experiment would have provided no 
information as we would have expected near exact agreement. Our theoretical values, 10.31(4) and 14.32(6), 
  are in perfect agreement with  values derived from the Stark shifts. 
  
  The values for the $5d-np$ and 
$6d-6p$ transitions are taken from the study of the inconsistencies in the lifetime and 
polarizability measurements in Cs \cite{safronova:2004}. The $5d-np$ values are scaled all-order values with the 
uncertainty defined as the spread of the \textit{ab initio} values containing the partial triple excitations
(SDpT) and scaled values based on both SD and SDpT calculations. The evaluation of the uncertainty of these values is described 
in \cite{safronova:2004}. The $6p-6d$ values are \textit{ab initio} SDpT results with the uncertainty defined as the 
difference of the \textit{ab initio} SD and SDpT calculations. We did not use the measured $5d$ lifetimes values 
\cite{5dtau} owing to the inconsistencies of these values with the measured Stark shifts \cite{tanner:1988, Stark2}. 
The uncertainties of the $5d_{3/2}-4f_{5/2}$ and $5d_{5/2}-4f_{7/2}$ matrix elements are 
estimated as the differences of the SD scaled results and \textit{ab initio} SDpT values that 
partially include triple excitations.  

  The SD all-order values for the $8s-np$ and $7d_j-nlj$ transitions were previously published in Refs.~\cite{mevan:2007,auzinsh:2007}, respectively. In summary, the uncertainties of our calculations are generally small, ranging from $0.2\%$ to about 1\%. The only exceptions are the transitions involving the $5d$ states 
  and some of the transition from the $6d$ states. We refer the reader to Ref.~\cite{safronova:2004} for 
  a detailed discussion of these transitions. We note that we may overestimate the uncertainty of our calculation for these transitions as our values for the $6p_{1/2}$ and $6p_{3/2}$ polarizabilities are in excellent agreement with the 
  experiment \cite{tanner:1988,Stark2}. As a result, the actual accuracy of our values of $5d$ polarizabilities may be actually 
  higher than we estimated.

 \begin{table*}
\caption{\label{tab7p1}  The contributions to the
scalar polarizability for the $7p_{1/2}$ state in cesium. The 
corresponding energy differences
and the absolute values of the lowest-order (DF) and final all-order electric-dipole
 reduced matrix elements are also listed. The energy differences are given in cm$^{-1}$.
 Electric-dipole  matrix elements  are given in atomic units ($ea_0$), and 
 polarizabilities are given in 10$^3$~$a_0^3$, where $a_0$ is Bohr radius. }
\begin{ruledtabular}
\begin{tabular}{lrrrrr}
\multicolumn{1}{c}{Contribution}&
\multicolumn{1}{c}{$\beta$}&
\multicolumn{1}{c}{$Z^{DF}_{\beta,7p_{1/2}}$ }&
\multicolumn{1}{c}{$Z^{SD}_{\beta,7p_{1/2}}$ }&
 \multicolumn{1}{c}{$E_{\beta}-E_{7p_{1/2}}$ }&
\multicolumn{1}{c}{$\alpha_0(7p_{1/2})$} \vspace{0.1cm}\\
\hline \hline \\[-0.4pc]
  $\alpha^{\text{main}}(ns)$        & $6s$      &   0.37 &   0.28  &-21765  &     -0.000           \\
                                    & $7s$      &  11.01 &  10.31  & -3230  &     -2.407(5)        \\
                                    & $8s$      &   9.53 &   9.31  &  2552  &      2.487(35)    \\
                                    & $9s$      &   2.04 &   1.97  &  5145  &      0.055           \\
                                    & $10s$     &   1.04 &   1.00  &  6535  &      0.011           \\
                                    & $11s$     &   0.68 &   0.65  &  7366  &      0.004           \\
                                    & $12s$     &   0.49 &   0.48  &  7904  &      0.002           \\
 $\alpha^{\text{tail}}(ns)$         &           &        &         &        &    0.012(12)    \\[0.5pc]
  $\alpha^{\text{main}}(nd_{3/2})$  & $5d_{3/2}$&   4.04 &   1.52  & -7266  &     -0.023        \\
                                    & $6d_{3/2}$&  19.62 &  17.99  &   824  &     28.74(70)     \\
                                    & $7d_{3/2}$&   4.03 &   6.56  &  4283  &      0.734(5)     \\
                                    & $8d_{3/2}$&   2.39 &   3.16  &  6046  &      0.121        \\
                                    & $9d_{3/2}$&   1.63 &   2.00  &  7063  &      0.042        \\
                                   & $10d_{3/2}$&   1.21 &   1.44  &  7703  &      0.020        \\
 $\alpha^{\text{tail}}(nd_{3/2})$  &            &        &         &        &    0.080(80)     \\[0.5pc]
  Total                            &            &        &         &        &    29.89(70)      \\
\end{tabular}
\end{ruledtabular}
\end{table*}

 \begin{table*}
\caption{\label{tab7p2}  The contributions to the
scalar and tensor polarizabilities  for the $7p_{3/2}$ state in cesium. The 
corresponding energy differences
and the absolute values of the lowest-order (DF) and final all-order electric-dipole
 reduced matrix elements are also listed. The energy differences are given in cm$^{-1}$.
 Electric-dipole  matrix elements  are given in atomic units ($ea_0$), and 
 polarizabilities are given in 10$^3$~$a_0^3$, where $a_0$ is Bohr radius. }
\begin{ruledtabular}
\begin{tabular}{lrrrrrr}
\multicolumn{1}{c}{Contribution}&
\multicolumn{1}{c}{$\beta$}&
\multicolumn{1}{c}{$Z^{DF}_{\beta,7p_{3/2}}$ }&
\multicolumn{1}{c}{$Z^{SD}_{\beta,7p_{3/2}}$ }&
 \multicolumn{1}{c}{$E_{\beta}-E_{7p_{3/2}}$ }&
\multicolumn{1}{c}{$\alpha_0(7p_{3/2})$} &
\multicolumn{1}{c}{$\alpha_2(7p_{3/2})$} \vspace{0.1cm}\\
\hline \hline \\[-0.4pc]
  $\alpha^{\text{main}}(ns)$        & $6s$      &  0.69   &   0.59&-21946  &   -0.001    &  0.001     \\
                                    & $7s$      & 15.35   &  14.32& -3411  &   -2.199(4) &  2.199(4)  \\
                                    & $8s$      & 14.28   &  14.07&  2371  &    3.05(3)  & -3.05(3)   \\
                                    & $9s$      &  2.86   &   2.73&  4964  &    0.055    & -0.055     \\
                                    & $10s$     &  1.44   &   1.38&  6354  &    0.011    & -0.011     \\
                                    & $11s$     &  0.93   &   0.89&  7185  &    0.004    & -0.004     \\
                                    & $12s$     &  0.68   &   0.65&  7722  &    0.002    & -0.002     \\
 $\alpha^{\text{tail}}(ns)$         &           &         &       &        &    0.01(1)  & -0.01(1)   \\[0.5pc]
  $\alpha^{\text{main}}(nd_{3/2})$  & $5d_{3/2}$&  1.69   &   0.58& -7447  &   -0.002    &  -0.0013   \\
                                    & $6d_{3/2}$&   8.86  &   8.07&  642   &   3.71(10)  &  2.97(8)    \\
                                    & $7d_{3/2}$&  2.11   &   3.32&  4102  &    0.098(1) &   0.079(1)  \\
                                    & $8d_{3/2}$&  1.19   &   1.54&  5865  &    0.015    &   0.012     \\
                                    & $9d_{3/2}$&  0.79   &   0.96&  6882  &    0.005    &   0.004     \\
                                   & $10d_{3/2}$&  0.58   &   0.68&  7522  &    0.002    &   0.002     \\
 $\alpha^{\text{tail}}(nd_{3/2})$  &            &         &       &        &    0.009(9) &   0.007(7)  \\[0.5pc]
   $\alpha^{\text{main}}(nd_{5/2})$ & $5d_{5/2}$&  5.02   &   1.87& -7350  &   -0.017    &    0.004    \\
                                    & $6d_{5/2}$& 26.61   &  24.35&   685  &   31.6(7)   &   -6.33(15) \\
                                    & $7d_{5/2}$&  6.30   &   9.64&  4122  &    0.825(6) &   -0.165(1)  \\
                                    & $8d_{5/2}$&  3.55   &   4.52&  5877  &    0.127    &   -0.025    \\
                                    & $9d_{5/2}$&  2.37   &   2.83&  6889  &    0.042    &   -0.009    \\
                                   & $10d_{5/2}$&  1.75   &   2.02&  7527  &    0.020    &   -0.004    \\
 $\alpha^{\text{tail}}(nd_{5/2})$  &            &         &       &        &    0.08(8)  &   -0.02(2)  \\[0.5pc]
 Total                             &            &         &       &        &   37.52(75) &  -4.41(17)  \\
\end{tabular}
\end{ruledtabular}
\end{table*}

As noted above, we used experimental energies for all of the main term calculations. Most of the energies 
values in this work are taken from the 1987 measurements by Weber and Sansonetti \cite{SW} and other 
values quoted in the same reference. The ionization potential value, required for the scaling 
procedure, is taken from the same work. The values of the several lower levels are taken from the NIST 
   \textit{Handbook of Basic Atomic Spectroscopic Data} \cite{NIST1}. The data for the $np_{3/2}$ levels are obtained by 
   combining the $np_{1/2}$ values  from \cite{SW} and fine-structure intervals
   from \cite{NIST2}. The data for the remaining few levels not given in either \cite{SW,NIST1}
   were taken from Ref.\cite{NIST2}. Since the energy denominators in the polarizability calculation are 
   small for some of the higher states (below 100~cm$^{-1}$), we compiled the list of the most accurate known
   energies. As a result, the polarizability values quoted in this work for the $7d$, $9d$, and $10d$
   states are  slightly different from the ones quoted in Ref.~\cite{auzinsh:2007} while the same matrix elements were
   used. We note that these differences are well within the uncertainties of the polarizability values. The uncertainties in the 
   values of the energies can be neglected in all cases.  
   
   Next,  we consider the examples of the polarizability calculation; one case is considered in detail for 
   each of the $nS$, $Np_{1/2}$, $Np_{3/2}$, $Nd_{3/2}$, and $Nd_{5/2}$ sequences of states.   We consider the following sample cases:   
$9s$, $7p_{1/2}$, $7p_{3/2}$, $8d_{3/2}$, and $8d_{5/2}$. 
   In addition, we consider the $5d_{3/2}$
   and $5d_{5/2}$ calculations separately as they do not follow the pattern of all other $Nd$ state 
   calculations. These are also the only cases where the tail contribution is significant
   and represent interesting exception among the states that we have considered.  
    \begin{table*}
    
\caption{\label{tab8d1}  The contributions to the
scalar and tensor polarizabilities for the $8d_{3/2}$ state in cesium. The 
corresponding energy differences
and the absolute values of the lowest-order (DF) and final all-order electric-dipole
 reduced matrix elements are also listed. The energy differences are given in cm$^{-1}$.
 Electric-dipole  matrix elements  are given in atomic units ($ea_0$), and 
 polarizabilities are given in 10$^3$~$a_0^3$, where $a_0$ is Bohr radius. }
\begin{ruledtabular}
\begin{tabular}{lrrrrrr}
\multicolumn{1}{c}{Contribution}&
\multicolumn{1}{c}{$\beta$}&
\multicolumn{1}{c}{$Z^{DF}_{\beta,8d_{3/2}}$ }&
\multicolumn{1}{c}{$Z^{SD}_{\beta,8d_{3/2}}$ }&
 \multicolumn{1}{c}{$E_{\beta}-E_{8d_{3/2}}$ }&
\multicolumn{1}{c}{$\alpha_0(8d_{3/2})$} &
\multicolumn{1}{c}{$\alpha_2(8d_{3/2})$ \vspace{0.1cm}}\\
\hline \hline \\[-0.4pc]
   $\alpha^{\text{main}}(np_{1/2})$ & $6p_{1/2}$ & 1.11 &   1.30 &   -16633   &    0.00    &   0.00       \\
                                    & $7p_{1/2}$ & 2.39 &   3.16 &    -6046   &   -0.06    &   0.06       \\
                                    & $8p_{1/2}$ & 5.55 &   9.18 &    -2102   &   -1.47(2) &   1.47(2)   \\
                                    & $9p_{1/2}$ &50.96 &  49.29 &     -174   &    -510(3) & 510(3)       \\
                                    & $10p_{1/2}$&20.43 &  14.02 &      916   &    7.85    &  -7.85       \\
                                    & $11p_{1/2}$& 5.84 &   4.50 &     1592   &    0.46    &  -0.46       \\
                                    & $12p_{1/2}$& 3.07 &   2.44 &     2041   &    0.11    &  -0.11       \\
 $\alpha^{\text{tail}}(np_{1/2})$   &            &      &        &            &    0.2(2)  &    -0.2(2)  \\[0.5pc]
  $\alpha^{\text{main}}(np_{3/2})$  & $7p_{3/2}$ & 1.19 &   1.54 &    -5865   &   -0.01    &    -0.01    \\
                                    & $8p_{3/2}$ & 2.97 &   4.71 &    -2020   &   -0.40    &    -0.32     \\
                                    & $9p_{3/2}$ &23.02 &  22.13 &     -130   & -138.3(9)  &  -110.6(7)  \\
                                    & $10p_{3/2}$& 8.43 &   5.53 &      942   &    1.18    &     0.95    \\
                                    & $11p_{3/2}$& 2.49 &   1.83 &     1610   &    0.08    &     0.06    \\
                                    & $12p_{3/2}$& 1.32 &   1.00 &     2053   &    0.02    &     0.01    \\
 $\alpha^{\text{tail}}(np_{3/2})$   &            &      &        &            &   0.04(4)  &    0.03(3)   \\[0.5pc]
  $\alpha^{\text{main}}(nf_{5/2})$  & $4f_{5/2}$ & 2.34 &   2.49 &    -3339 &     -0.07    &   0.01     \\
                                    & $5f_{5/2}$ &19.18 &  26.06 &     -840 &    -29.6(4)  &   5.9(1)  \\
                                    & $6f_{5/2}$ &70.91 &  65.22 &      518 &    300(3)    & -60.0(6)   \\
                                    & $7f_{5/2}$ & 8.74 &   0.33 &     1337 &      0.00    &   0.00     \\
                                    & $8f_{5/2}$ & 5.71 &   1.28 &     1868 &      0.03    &  -0.01     \\
 $ \alpha^{\text{tail}}(nf_{5/2})$  &            &      &        &            &      1(1)  &   -0.2(2)   \\[0.5pc]
  Total                             &            &      &        &            &   -369(5)  &      339(4)   \\
\end{tabular}
\end{ruledtabular}
\end{table*}

 \begin{table*}
\caption{\label{tab8d2}  The contributions to the
scalar and tensor polarizabilities for the $8d_{5/2}$ state in cesium. The 
corresponding energy differences
and the absolute values of the lowest-order (DF) and final all-order electric-dipole
 reduced matrix elements are also listed. The energy differences are given in cm$^{-1}$.
 Electric-dipole  matrix elements  are given in atomic units ($ea_0$), and 
 polarizabilities are given in 10$^3$~$a_0^3$. }
\begin{ruledtabular}
\begin{tabular}{lrrrrrr}\\[-0.4pc]
\multicolumn{1}{c}{Contribution}&
\multicolumn{1}{c}{$\beta$}&
\multicolumn{1}{c}{$Z^{DF}_{\beta,8d_{5/2}}$ }&
\multicolumn{1}{c}{$Z^{SD}_{\beta,8d_{5/2}}$ }&
 \multicolumn{1}{c}{$E_{\beta}-E_{8d_{5/2}}$ }&
\multicolumn{1}{c}{$\alpha_0(8d_{5/2})$} &
\multicolumn{1}{c}{$\alpha_2(8d_{5/2})$ \vspace{0.1cm}}\\
\hline \\[-0.4pc]
  $\alpha^{\text{main}}(nP_{3/2})$  & $6p_{3/2}$ &   1.59  & 1.81 & -16091   &    0.00     &    0.00     \\
                                    & $7p_{3/2}$ &   3.55  & 4.52 &  -5877   &   -0.08     &    0.08     \\
                                    & $8p_{3/2}$ &   8.82  &13.65 &  -2031   &   -2.24(2)  &    2.24(2)  \\
                                    & $9p_{3/2}$ &  69.07  &66.57 &   -141   &   -765(5)   &  765(5)     \\
                                    & $10P_{3/2}$&  25.43  &17.30 &    931   &    7.84     &   -7.84     \\
                                    & $11P_{3/2}$&   7.51  & 5.68 &   1598   &    0.49     &   -0.49     \\
                                    & $12P_{3/2}$&   3.98  & 3.10 &   2041   &    0.11     &   -0.11      \\
 $\alpha^{\text{tail}}(nP_{3/2})$   &            &         &      &          &    0.2(2)   &   -0.2(2)   \\[0.5pc]
  $\alpha^{\text{main}}(nF_{5/2})$  & $4f_{5/2}$ &   0.62  & 0.67 &  -3350.4 &      0.00   &     0.00    \\
                                    & $5f_{5/2}$ &   5.11  & 6.85 &   -851.3 &     -1.34(2)&    -1.53(2) \\
                                    & $6f_{5/2}$ &  18.97  &17.54 &    506.6 &     14.8(1) &    16.9(2) \\
                                    & $7f_{5/2}$ &   2.37  & 0.04 &   1325.1 &      0.00   &     0.00   \\
                                    & $8f_{5/2}$ &   1.54  & 0.42 &   1855.9 &      0.00   &     0.00    \\
 $\alpha^{\text{tail}}(nF_{5/2})$   &            &         &      &          &     0.05(5) &    0.05(5)   \\[0.5pc]
  $\alpha^{\text{main}}(nF_{7/2})$  & $4f_{7/2}$ &   2.79  & 2.99 &  -3350.7 &     -0.06   &      0.02   \\
                                    & $5f_{7/2}$ &  22.82  &30.60 &   -851.6 &    -26.8(4) &      9.6(1)   \\
                                    & $6f_{7/2}$ &  84.82  &78.43 &    506.5 &    296(3)   &     -106(1)   \\
                                    & $7f_{7/2}$ &  10.61  & 0.19 &   1325.0 &      0.00   &      0.00     \\
                                    & $8f_{7/2}$ &   6.89  & 1.86 &   1855.8 &      0.05(5)&     -0.02     \\
 $\alpha^{\text{tail}}(nF_{7/2})$   &            &         &      &          &     1(1)    &    -0.4(4)  \\[0.5pc]
  Total                             &            &         &      &          &    -475(5)  &    678(5)   \\
\end{tabular}
\end{ruledtabular}
\end{table*}

     We consider the $9s$ case first. The detailed breakdown of the $9s$ 
     polarizability calculation is given in Table~\ref{tab9s}. Each contribution to the main term, i.e. the contributions from the 
     $6p, 7p, 8p, 9p, 10p, 11p$ and $12p$ states are given separately, and the tail 
     terms are grouped together for the $np_{1/2}$ and $np_{3/2}$ contributions.   The 
corresponding main term energy differences
and the absolute values of the lowest-order (DF) and final all-order electric-dipole
 reduced matrix elements are also listed. The lowest-order values are given to illustrate the size of the 
 correlation corrections for these transitions. The energy differences are given in cm$^{-1}$.
 Electric-dipole  matrix elements  are given in atomic units ($ea_0$), and 
 polarizabilities are given in 10$^3$~$a_0^3$. The core contribution is negligible in this 
 case (0.015 in the units of Table~\ref{tab9s}) and is not listed.  We find that two of the transitions,
 $9s-9p_{1/2}$ and $9s-9p_{3/2}$, give dominant contributions while two other,  $9s-8p_{1/2}$ and $9s-8p_{3/2}$,
 are large and have to be calculated accurately. We note that there is rather significant cancellation between the 
 $9s-9p_{j}$ and $9s-8p_{j}$ contributions.  The dominant contribution is this case may 
 have been easily predicted simply based on the size of the energy intervals listed in the 
 fifth column of the table. We also find that  all other contributions 
 with the exception of the $9s-10p_{1/2}$ and $9s-10p_{3/2}$ contribution
 are very small and may be simply omitted 
 without the loss of accuracy. The main uncertainty comes from the uncertainty in the $9s-9p_{3/2}$
 transition. The precision our calculation in this case is expected to be very high as the 
 correlation correction is small as illustrated by the comparison of the lowest-order and final values of the electric-dipole matrix elements. The final uncertainty is evaluated to be 0.7\%. The breakdown of the calculation of the 
 other $Ns$ polarizabilites considered in this work is similar to the one for the $9s$ state with the 
 exception of the $6s$ state. For all other cases, the dominant contributions come from the 
 $Ns - Np_{1/2}$ and the $Ns - Np_{3/2}$ matrix elements, while the other important contributions
 come from the $Ns - (N-1)p_{1/2}$ and the $Ns - (N-1)p_{3/2}$ matrix elements. 
  The polarizability of the $6s$ state is overwhelmingly dominated by the 
 contribution of the $6s - 6p_{1/2}$ and the $6s - 6p_{3/2}$ transitions. These two transitions add coherently and 
 account for the 96\% of the total value. The calculation of the $8s$ polarizability is described in detail 
 in Ref.~\cite{mevan:2007}. We limited this work by the 
 $12s$ state as the $13p_{j}$ states needed for the calculation of the $13s$ polarizability 
 do not quite fit inside of our cavity and the basis set energies of the $13p$ states deviate 
 from the DF energies.      
  
  The breakdown of the contributions to the $7p_{1/2}$ and $7p_{3/2}$ polarizabilities is given in 
  Tables~\ref{tab7p1} and \ref{tab7p2}, respectively. All tables illustrating the 
  contributions to polarizabilities are structured in the same way. 
   In the case of the $7p_{1/2}$ polarizability,
  the dominant contribution comes from a single transitions, $7p_{1/2}-6d_{3/2}$, as none of the other levels 
  are as close to the $7p_{1/2}$ levels as the $6d_{3/2}$ level. The contribution from the next transition,
   $7p_{1/2}-7d_{3/2}$, is significantly smaller, only 2\% of the dominant contribution. Interestingly, 
   the contributions of the $7p_{1/2}-7s$ and $7p_{1/2}-8s$ transitions, while  being 10\% of the 
   main contribution, cancel out nearly exactly. We note that while significant cancellation is present 
   for all other $Np_{1/2}$ cases, it is the most severe in the case of the $7p_{1/2}$ state.
   The  tail contribution is larger than for the 
   $Ns$ calculation but is still very small, 0.3\%.  We assume 100\% uncertainty in the tail contributions
    in all of our calculations for consistency. It is still negligible for all of the cases with the exception of the $5d$ calculation. 
    \begin{table*}
\caption{\label{tab5d1}  The contributions to the
scalar and tensor polarizabilities for the $5d_{3/2}$ and $5d_{5/2}$ states n cesium. The 
corresponding energy differences
and the absolute values of the lowest-order (DF) and final all-order electric-dipole
 reduced matrix elements are also listed. The energy differences are given in cm$^{-1}$.
 Electric-dipole  matrix elements  are given in atomic units ($ea_0$), and 
 polarizabilities are given in 10$^3$~$a_0^3$, where $a_0$ is Bohr radius. }
\begin{ruledtabular}
\begin{tabular}{lrrrrrr}
\multicolumn{1}{c}{Contribution}&
\multicolumn{1}{c}{$\beta$}&
\multicolumn{1}{c}{$Z^{DF}_{\beta,5d_{3/2}}$ }&
\multicolumn{1}{c}{$Z^{SD}_{\beta,5d_{3/2}}$ }&
 \multicolumn{1}{c}{$E_{\beta}-E_{5d_{3/2}}$ }&
\multicolumn{1}{c}{$\alpha_0(5d_{3/2})$} &
\multicolumn{1}{c}{$\alpha_2(5d_{3/2})$ \vspace{0.1cm}}\\
\hline \hline \\[-0.4pc]
  $\alpha^{\text{main}}(np_{1/2})$  & $6p_{1/2}$ & 8.98 & 7.06 & -3321  &   -0.550(24)&    0.550(24)    \\
                                  & $7p_{1/2}$ & 4.04 &  1.52&  7266  &    0.012    &    -0.012       \\
$\alpha^{\text{tail}}(np_{1/2})$  &            &      &      &        &      0.002  &      -0.002   \\[0.5pc]
  $\alpha^{\text{main}}(np_{3/2})$  & $6p_{3/2}$ & 4.06 & 3.19 & -2767  &   -0.134(6) &   -0.107(5)     \\ 
                                    & $7p_{3/2}$ & 1.69 & 0.58 &  7447  &    0.002    &    0.001      \\
$\alpha^{\text{tail}}(np_{3/2})$    &            &      &      &        &    0.000    &    0.000   \\[0.5pc]
  $\alpha^{\text{main}}(nf_{5/2})$  & $4f_{5/2}$ &10.66 & 7.11 &  9973  &    0.186(27)&    0.037(5)   \\
                                    & $5f_{5/2}$ & 4.72 & 3.34 & 12472  &    0.033    &   -0.007       \\
                                    & $6f_{5/2}$ & 2.90 & 2.24 & 13830  &    0.013    &   -0.003       \\
                                    & $7f_{5/2}$ & 2.04 & 1.66 & 14649  &    0.007    &   -0.001       \\
                                    & $8f_{5/2}$ & 1.55 & 1.30 & 15180  &    0.004    &   -0.001       \\
 $ \alpha^{\text{tail}}(nf_{5/2})$  &            &      &      &        &    0.059(59)&   -0.012(12)  \\[0.5pc]
  Total                             &            &      &      &        &   -0.352(69)&    0.370(28)   \\
\hline\\[-0.4pc]
\multicolumn{1}{c}{Contribution}&
\multicolumn{1}{c}{$\beta$}&
\multicolumn{1}{c}{$Z^{DF}_{\beta,5d_{5/2}}$ }&
\multicolumn{1}{c}{$Z^{SD}_{\beta,5d_{5/2}}$ }&
 \multicolumn{1}{c}{$E_{\beta}-E_{5d_{5/2}}$ }&
\multicolumn{1}{c}{$\alpha_0(5d_{5/2})$} &
\multicolumn{1}{c}{$\alpha_2(5d_{5/2})$ \vspace{0.1cm}}\\
\hline  \\[-0.4pc]
  $\alpha^{\text{main}}(np_{3/2})$  & $6p_{3/2}$ & 12.19 &  9.66 & -2865&     -0.794(33)&    0.794(33)     \\   
                                    & $7p_{3/2}$ &  5.02 &  1.87 &  7350&      0.012    &   -0.012        \\
$\alpha^{\text{tail}}(np_{3/2})$  &              &       &       &      &      0.002    &   -0.002 \\[0.5pc]
  $\alpha^{\text{main}}(nf_{5/2})$  & $4f_{5/2}$ &  2.84 &  1.93 &  9876&      0.009(3) &    0.011(3)    \\ 
                                    & $4f_{5/2}$ &  1.26 &  0.91 & 12375&      0.002    &    0.002     \\
$\alpha^{\text{tail}}(nf_{5/2})$    &            &       &       &      &      0.004(3) &    0.004(3) \\[0.5pc]
  $\alpha^{\text{main}}(nf_{7/2})$  & $4f_{7/2}$ &112.70 &  8.62 &  9875&      0.184(24)&    -0.066(9)  \\
                                    & $5f_{7/2}$ &  5.64 &  4.08 & 12375&      0.033    &    -0.012      \\
                                    & $6f_{7/2}$ &  3.46 &  2.73 & 13733&      0.013    &    -0.005      \\
                                    & $7f_{7/2}$ &  2.44 &  2.01 & 14551&      0.007    &    -0.002      \\
                                    & $8f_{7/2}$ &  1.86 &  1.57 & 15082&      0.004    &    -0.001      \\
 $ \alpha^{\text{tail}}(nf_{7/2})$  &            &       &       &      &    0.056(56)  &  -0.020(20) \\[0.5pc]
  Total                             &            &       &       &      &   -0.453(70)  &    0.691(40)   \\ 
\end{tabular}
\end{ruledtabular}
\end{table*}

   As noted above, there are three types of the transitions contributing to the polarizabilities of the $np_{3/2}$
   states. The dominant contribution comes from the single transition as in the case of the $7p_{1/2}$
   polarizabilites, $7p_{3/2}-6d_{5/2}$. The contribution of the $7p_{3/2}-6d_{3/2}$ transition is 
   10 times as small as the dominant one. Again, the contributions from the $7p_{3/2}-7s$ and $7p_{3/2}-8s$
  partially cancel, but the cancellation is not as complete as in the case of the $7p_{1/2}$ states. 
  
  While the calculations of the scalar and tensor polarizabilities use the 
  same matrix elements and energies and only differ by the angular factors, the uncertainty of the 
  $7p_{3/2}$ tensor polarizability calculation (4\%) is twice as high as that of the scalar polarizability
  owing to the significant cancellation of the terms contributing to the tensor polarizability. 
  The relative accuracy of the calculation of the tensor polarizability calculation gradually improves to 1\% for the $10p_{3/2}$
  state but this uncertainty  is still more than twice as high as the uncertainty of the corresponding 
  scalar polarizability calculation (0.4\%).    The breakdown of all other $np_{1/2}$ and $np_{3/2}$ polarizabilites parallels the one of the $7p_{1/2}$
   and $7p_{3/2}$ states. 
 
  The contributions to
scalar and tensor polarizabilities for the $8d_{3/2}$ and $8d_{5/2}$ states in cesium are given by Tables~\ref{tab8d1}
and \ref{tab8d2}. For the $8d_{3/2}$ states, three contributions are dominant, $8d_{3/2}-9p_{1/2}$, $8d_{3/2}-9p_{1/2}$,
and $8d_{3/2}-6f_{5/2}$ for both scalar and tensor polarizabilites. Unlike the case of the $Np_{3/2}$ 
states, significant cancellations are observed between terms for both scalar and tensor polarizabilites. 
We would like to specifically note interesting problem with the $8d_{3/2}-7f_{5/2}$ transition. While the DF 
value for the transition is 8.74,  the final all-order number is very small, 0.33 owing to extremely 
large correlation correction that essentially cancels the lowest order. We also note that the \textit{ab initio}
all-order value for this transition (0.73) significantly differs from the scaled values. While we assigned 
this value 100\% uncertainty, the resulting uncertainty in the polarizability value
is negligible.  

 We observe similar problem with the $8d_{5/2}-7f_{7/2}$ transition as well 
as similar transitions for other values of $N$ and $n$ with the exception of the $5d-4f$ transitions. For the 
case of the $6d-5f$ transition, the cancellation of the lowest order and the 
correlation correction is less severe.  We note that the correlation correction to the 
previous transition in the sequence, $8d_{3/2}-6f_{5/2}$ is small, only 8\%. Similar issue exists for the 
next in line transition, $8d_{3/2}-8f_{5/2}$, but its contribution was too small to warrant its more accurate consideration. 
 
 The $8d_{3/2}-7f_{5/2}$ and $8d_{5/2}-7f_{7/2}$ transitions are  two
  of the very few transitions for which we conducted the scaling but did not list 
the values in the Table~\ref{me} of the recommended matrix elements as the uncertainties of these values are
 very high. In general, 
if the main term transition was not listed in Table~\ref{me}, we used \textit{ab initio} SD value and did not conduct the evaluation of the uncertainty. The contributions of these terms are small enough so their contribution to the 
total uncertainties would be negligible.  Again, significant cancellations are observed between the terms. The polarizability calculation of the all other $Nd$ is similar to the $8d$ examples with the 
exception of the $5d$ scalar polarizability calculation, which is anomalous and is discussed separately below. The calculation
 of the $7d$, $9d$, and $10d$ polarizabilities was discussed in detail in Ref.~\cite{auzinsh:2007}.  
  \begin{table*}
\caption{\label{comp1} Comparison of the Cs scalar polarizabilities with 
other theory and experiment. All values are given in 10$^3$~$a_0^3$.
$^a$Recommended value from Ref.~\cite{derevianko:1999}, 
$^b$\textit{ab initio} all-order value from Ref.~\cite{safronova:1999}, 
$^c$Ref.\cite{patil:1997}, 
$^d$Ref.\cite{amini:2003}, 
$^e$derived from the  Ref.~\cite{new7s} $7s-6s$ Stark shift measurement and the $6s$ result from \cite{amini:2003}, 
$^f$Ref.~\cite{mevan:2007},
$^g$Refs.~\cite{WW,wijngaarden:1994},
$^h$derived from Ref.~\cite{Stark2} D1 line
Stark shift measurement and the $6s$ result from \cite{amini:2003},
$^i$Ref.~\cite{domelunksen:1983},
$^j$derived from Ref.~\cite{tanner:1988} D2 line
Stark shift measurement and the $6s$ result from \cite{amini:2003},
$^k$Ref.~\cite{khvoshtenko:1968},
$^l$Ref.~\cite{wessel:1987},
$^m$Ref.~\cite{fredriksson:1977}
$^n$Ref.~\cite{xia:1997}. 
}
\begin{ruledtabular}
\begin{tabular}{lllllll}
State   &  $6s$       & $ 7s$       & $8s$       &   $ 9s$      &  $10s$       & $11s$      \\
\hline \\[-0.4pc]
Present &  0.3984(7)  &  6.238(41)  &  38.27(28) &  153.7(1.0)  &  478(3)      & 1246(8)  \\       
Ref.~\cite{wijngaarden:1994}  &  0.394      &  6.14       &  37.9      &  153         &   475        & 1240     \\
Theory  &  0.3999(19)$^a$& 6.272$^b$&            &              &            &   \\
        &  0.4091$^c$ &             &            &              &             &          \\
 Expt.  &  0.4010(6)$^d$& 6.238(6)$^e$& 38.06(25)$^f$&          &   479(1)$^g$   &  1246(1)$^g$\\ [0.5pc]
 \hline
State   &   $6p_{1/2}$&  $7p_{1/2}$ &  $8p_{1/2}$&   $9p_{1/2}$ &  $10p_{1/2}$& $12s $\\[0.2pc]
\hline
Present &    1.338(54)&    29.9(7)  &    223(2)  &     1021(7)  &    3499(19) & 2866(30) \\
Ref.~\cite{wijngaarden:1994}  &    1.29     &    29.4     &    221     &     1020     &    3490& 2840\\
 Expt.  &    1.3284(6)$^h$&29.6(6)$^i$&            &              &           &2871(2)$^g$ \\ [0.5pc]
 \hline
State   &  $6p_{3/2}$ &  $7p_{3/2}$ &  $8p_{3/2}$&    $9p_{3/2}$&  $10p_{3/2}$& \\[0.2pc]
\hline
Present &  1.648(56)  &     37.5(8) &     284(3) &     1312(7)  &   4522(19)  &\\
Ref.~\cite{wijngaarden:1994}  &  1.60       &     36.9    &     282    &     1310     &   4510&\\
 Expt.  &  1.641(2)$^j$&  37.9(8)$^k$&          &               &                            &       \\ [0.5pc]
 \hline
 State   &  $5d_{3/2}$ &  $6d_{3/2}$ &  $7d_{3/2}$&   $8d_{3/2}$ &   $9d_{3/2}$ &  $10d_{3/2}$ \\ [0.2pc]
 \hline 
Present &  -0.352(69) &   -5.68(45) &   -66.7(1.7)&   -369(5)   &  -1402(13)   &   -4234(32)  \\  
Ref.~\cite{wijngaarden:1994}  &  -0.418     &   -5.32     &   -65.2    &    -366      &  -1400       &   -4220\\
 Expt.  &             &             &  -60(8)$^l$&              &   -1450(120)$^m$& -4185(4)$^n$\\  [0.5pc]
 \hline
State   &  $5d_{5/2}$ &  $6d_{5/2}$ &  $7d_{5/2}$&   $8d_{5/2}$ &   $9d_{5/2}$ &  $10d_{5/2}$ \\ [0.2pc] 
\hline
Present &   -0.453(70) &   -8.37(55) &  -88.8(2.0)&    -475(5)   &   -1777(14)  &   -5316(38)  \\  
Ref.~\cite{wijngaarden:1994}  &   -0.518    &   -7.95     &  -87.1     &    -472      &   -1770      &   -5300\\
 Expt.  &             &             &   -76(8)$^l$&             & -2050(100)$^m$&   -5303(8)$^n$ \\    
\end{tabular}
\end{ruledtabular}
\end{table*}

 \begin{table*}
\caption{\label{comp2} Comparison of the Cs tensor polarizabilities with 
other theory and experiment. All values are given in 10$^3$~$a_0^3$. 
$^a$Ref.~\cite{tanner:1988},
$^b$Ref.~\cite{khvoshtenko:1968},
$^c$Ref.~\cite{fredriksson:1977}
$^d$Ref.~\cite{lurio}
$^e$Ref.~\cite{domelunksen:1983},
$^f$Ref.~\cite{auzinsh:2006},
$^g$Ref.~\cite{xia:1997}
$^h$Ref.~\cite{wessel:1987}
}
\begin{ruledtabular}
\begin{tabular}{lllllll}
State    &  $6p_{3/2}$ &   $7p_{3/2}$ &   $8p_{3/2}$ &    $9p_{3/2}$&     $10p_{3/2}$&  \\
\hline \\[-0.4pc]
Present  &  -0.261(13) &    -4.41(17) &     -30.6(6) &     -135(2)  &     -451(5)    & \\
Ref.~\cite{wijngaarden:1994}   &  -0.223     &    -4.28     &     -30.2    &     -134     &     -449 & \\
Expt.    & -0.2624(15)$^a$&  -4.43(12)$^b$& -30.7(1.2)$^c$ & & &\\ 
          &             &  -4.33(17)$^d$& & & &\\ 
         &              &  -4.00(8)$^e$ & & & & \\   [0.5pc]
         \hline
State    &  $5d_{3/2}$ &   $6d_{3/2}$ &   $7d_{3/2}$ &   $8d_{3/2}$ &    $9d_{3/2}$  &  $10d_{3/2}$\\[0.2pc] 
\hline
Present  &   0.370(28)&    8.77(36)  &   71.1(1.2)  &    339(4)    &     1189(10)   &   3416(26) \\
Ref.~\cite{wijngaarden:1994}   &   0.380     &    8.62      &   70.4       &    336       &     1190       &    3410       \\
Expt.    &             &              &  74.5(2.0)$^f$&   333(16)$^c$&   1183(35)$^f$& 3401(4)$^g$       \\ [0.5pc]
\hline
State    &  $5d_{5/2}$ &   $6d_{5/2}$ &   $7d_{5/2}$ &   $8d_{5/2}$ &    $9d_{5/2}$  &  $10d_{5/2}$ \\ [0.2pc]
\hline 
Present  &   0.691(40) &    17.33(50) &     142(2)   &       678(5) &      2386(13)  &    6869(34) \\  
Ref.~\cite{wijngaarden:1994}   &  0.703      &    17.00     &     140      &       675    &      2380  &   6850  \\
Expt.    &            &              &   129(4)$^h$ &    734(4)$^c$ &     2660(140)$^c$ & 6815(20)$^g$ \\
       &             &              &                &              &                 & 7140(36)$^c$              \\
\end{tabular}
\end{ruledtabular}
\end{table*}

 The contributions to the
scalar and tensor polarizabilities for the $5d_{3/2}$ and $5d_{5/2}$ states in cesium are given in Table~\ref{tab5d1}. 
    We grouped small contributions of the $5d-np_j$ and $5d-nf_{5/2}$ transitions together with the tail in this table. 
   Comparison of the $5d_{3/2}$ and $8d_{3/2}$ tables (as well as all the other $nd_{3/2}$ contribution
   breakdowns) shows the  $5d_{3/2}$ scalar polarizability case to be anomalous. In this case, none of the $5d-nf$  energy denominators are small, and the largest contribution from $nf_{5/2}$ states is still a third of the 
   one from the dominant $5d_{3/2}-6p_{1/2}$ transition. There is also no damping of the remaining
   $5d_{3/2}-nf_{5/2}$ contributions  observed for the higher $8d_{3/2}-nf_{5/2}$ transitions. Therefore, there is 
   basis to assume that the DF tail is substantially overestimated. 
   It may be overestimated by about 
   15-20\% based on the comparison of the DF and the all-order matrix element values. As a result,
   the tail contribution is 25\% of the total contribution of the $5d_{3/2}-nf_{5/2}$ sum and  
    its uncertainty gives the 
   dominant contribution to the uncertainty of the $5d_{3/2}$ scalar polarizability. We note that the $5d_{3/2}-nf_{5/2}$ tensor 
   polarizability tail is small with comparison to the dominant  $5d_{3/2}-6p_{1/2}$ contribution, and  
   its contribution to the total uncertainty is small. As a result, the $5d_{3/2}$ 
   tensor polarizability calculation is similar to the  $8d_{3/2}$ one. Its reduced accuracy is due to 
   much larger correlation correction to the $5d_{3/2}-6p_{1/2}$ matrix element
   in comparison to the $8d_{3/2}-9p_{1/2}$ one as illustrated by the comparison of the lowest-order and the all-order $5d_{3/2}-6p_{1/2}$ and 
   $8d_{3/2}-9p_{1/2}$ data. 
  The analysis of the $5d_{5/2}$ polarizability is similar to that of the $5d_{3/2}$ one. The main contribution to the 
  uncertainty of the scalar polarizability comes from the $5d_{5/2}-nf_{7/2}$ tail and the uncertainties of the 
  dominant terms are substantially larger than the uncertainties for the other $Nd$ states for both 
  scalar and tensor polarizabilites owing to large correlation 
  correction of the corresponding transitions.  
 
 \section{Comparison with other theory and experiment}
 Our final results for the scalar and tensor Cs polarizabilities  and their uncertainties are compared with 
 other theoretical and experimental values in 
 Tables~\ref{comp1} and \ref{comp2}, respectively. As we noted above, the theory values for the
  $8s, 7d, 9d$, and $10d$ polarizabilites  from Refs.~\cite{mevan:2007,
  auzinsh:2007} differ very slightly from the present values since they are obtained using the 
  same values of the matrix elements but more accurate energies. Therefore, we do not 
  quote theory values from Refs.~\cite{mevan:2007,auzinsh:2007} separately in Tables~\ref{comp1} and \ref{comp2}.
   The experimental values for the $7s$, $6p_{1/2}$, and 
 $6p_{3/2}$ states are obtained by combining the most accurate measurements of the $7s-6s$ \cite{new7s}, 
 $6p_{1/2}-6s$ \cite{Stark2}, and $6p_{3/2}-6s$ \cite{tanner:1988} Stark shifts with the recent measurement of the $6s$ polarizability \cite{amini:2003},
 respectively.  
 We find excellent agreement of our values with high-precision 
 measurements of Refs.~\cite{amini:2003,Stark2,WW,xia:1997,tanner:1988,mevan:2007,auzinsh:2006}. 
 Disagreements with older values for the $Nd$ states  are discussed in detail in Ref.~\cite{auzinsh:2007}.
 In all cases where the new measurements are available, our data support most precise 
 measurements. In particular, we find that our method works very well for even such 
highly-excited states as $12s$ and $10d$. 
  
We also compare our values with the van Wijngaarden and Li \cite{wijngaarden:1994} work where
 the extensive calculations of the polarizabilities of 
cesium $Ns$, $Np$, $Nd$, and $Nf$ states were carried out using the Coulomb approximation. Our values are in excellent 
agreement with those results for higher excited states where the method of Ref.~\cite{wijngaarden:1994}
is expected to work well. 

\section{Conclusion}

We have carried out a systematic study of the Cs electric-dipole static polarizabilities for the $Ns$ ($N = 6 - 12$), 
$Np_j$ ($N = 6 - 10$), and $Nd_j$ ($N = 5 - 10$) states using the relativistic all-order method. 
The recommended values for the polarizabilites of all these states are given and their
 uncertainties are estimated. This work involved the calculation of 317 electric-dipole
transition in Cs. Recommended values for the 91 transitions that give the dominant contributions to the 
polarizabilities are presented together with their uncertainties. 
Our polarizability values are compared with other theory and experiment. Our data are found to be in 
excellent agreement with the high-precision measurements.  These calculations provide a theoretical
 benchmark for a large number of Cs electric-dipole matrix elements and polarizabilities.
\begin{acknowledgments}
 The work of EIT and MSS  was
supported in part by National Science Foundation Grant  No.\
PHY-04-57078. 
\end{acknowledgments}

\end{document}